\documentclass[aps,prl,reprint,letterpaper,showpacs]{revtex4-1}
\usepackage{amssymb}
\usepackage{keyval}%
\usepackage{graphicx}
\usepackage{dcolumn}
\usepackage{bm}
\usepackage{color}
\usepackage{fancyhdr}

\newcommand{\ignore}[1]{}

\newcommand{\material}{Sr$_3$CuIrO$_6$}

\begin{document}
\title{Spin Frustration and a `Half Fire, Half Ice' Critical Point from Nonuniform $g$-Factors}
\author{Wei-Guo Yin}
\email{wyin@bnl.gov}
\author{Christopher R. Roth}
\author{Alexei M. Tsvelik}
\affiliation{Condensed Matter Physics and Materials Science Division, Brookhaven National Laboratory, Upton, New York 11973, USA}
\date{\today}


\begin{abstract}
It is demonstrated that novel spin frustration can be induced in ferromagnets with nonuniform Land\'{e} $g$-factors. The frustrated state is characterized by a mutual interplay of typical ferromagnetic (FM) and antiferromagnetic (AF) features, such as the zero-field susceptibility being FM-like at low temperatures but AF-like at high temperatures. It is also found to contain an exotic zero-temperature `half fire, half ice' critical point at which the spins on one sublattice are fully disordered and on the other one are fully ordered. We suggest that such frustration may occur in a number of copper-iridium oxides such as Sr$_3$CuIrO$_6$. We also anticipate a realization of the frustration and `partial fire, partial ice' states in certain antiferromagnets, lattice gas, and neuron systems.
\end{abstract}

\maketitle
\begin{figure}[b]
        \includegraphics[width=0.9\columnwidth,clip=true,angle=0]{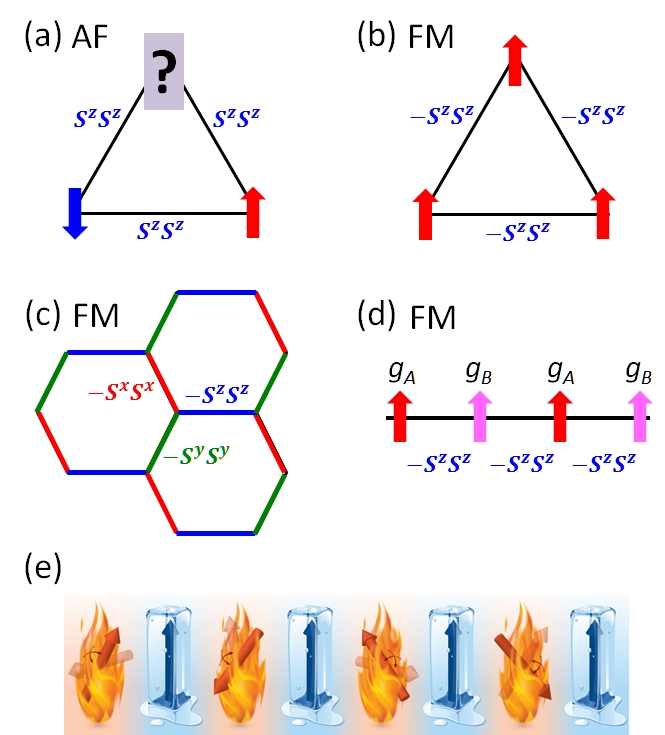}
        \caption{
        (a) Frustrated spins on triangular lattice with AF interactions. (b) Unfrustrated spins for FM interactions. (c) Frustrated spins on honeycomb lattice with FM interactions whose anisotropic axes are bond dependent \cite{Kitaev2006,Chaloupka2010}. (d) Spins in a chain with uniform FM interactions and alternating $g$-factors ($-g_B>g_A>0$) are found frustrated. (e) A cartoon  illustration of the `half-fire, half-ice' critical point at which the spins on one sublattice are  fully disordered and on the other are  fully ordered.
        \label{geometry}}
\end{figure}

Frustrated magnets are known to give rise to exotic magnetic states such as spin ice, spin glass, and spin liquid that may play important roles in quantum computing, spintronics, and unconventional superconductivity \cite{Balents2010,Mila2011}. The essence of frustration is known to be ground-state degeneracy which emerges as a consequence of competing exchange interactions among the spins. Such frustration demands that some or all of the exchange interactions be antiferromagnetic (AF) [Figs.~\ref{geometry}(a)(b)],
or in case the exchange interactions are all FM, that they have strong direction-dependent anisotropy [Fig.~\ref{geometry}(c)].

Here we suggest a different mechanism of frustration, the one which can exist in magnets with uniform FM exchange interactions and is related to nonuniformity of the Land\'{e} $g$-factors. It is likely to exist in a class of copper-iridium oxides such as (Sr,Ba)$_{2+n}$CuIr$_n$O$_{3n+3}$ ($n$=1$-$4), Ba$_9$Cu$_2$Ir$_5$O$_{21}$, Ba$_{14}$Cu$_3$Ir$_8$O$_{33}$, and Ba$_{16}$Cu$_3$Ir$_{10}$O$_{39}$ \cite{Niazi2002,Battle1997,Battle1998,Blake1998}. For example, Sr$_3$CuIrO$_6$, once believed to be a model isotropic spin one-half ($S=\frac{1}{2}$) chain ferromagnet \cite{Nguyen1996,Furusaki1994}, was found to display a strange spin glassy behavior with a typical FM and AF zero-field susceptibility at low and high temperatures, respectively \cite{Niazi2002}. These observations  seemed to suggest the existence of substantial AF exchange interactions  \cite{Niazi2002,Battle1997,Battle1998,Blake1998}, but the  recent first-principles symmetry analysis and resonant inelastic X-ray scattering experiments \cite{Yin2013,Yin2012} have reestablished the uniform FM nature of the effective exchange interactions in Sr$_3$CuIrO$_6$, though also demonstrating a strong Ising-like anisotropy [Fig.~\ref{geometry}(d)]. Most importantly, it was found \cite{Yin2013} that the different combinations of spin-orbit coupling and crystal-field splitting for the Cu and Ir ions make the $g$-factors 
strongly site dependent, viz. $g \approx 2$ and $-3$ for the Cu and Ir sites, respectively, along the easy axis (see Supplemental Material \cite{SI}). Below we demonstrate that such disparity in the $g$-factors does generate strong frustration in both zero-field and critical-field limits. 

\begin{figure*}[t]
        \includegraphics[width=2\columnwidth,clip=true,angle=0]{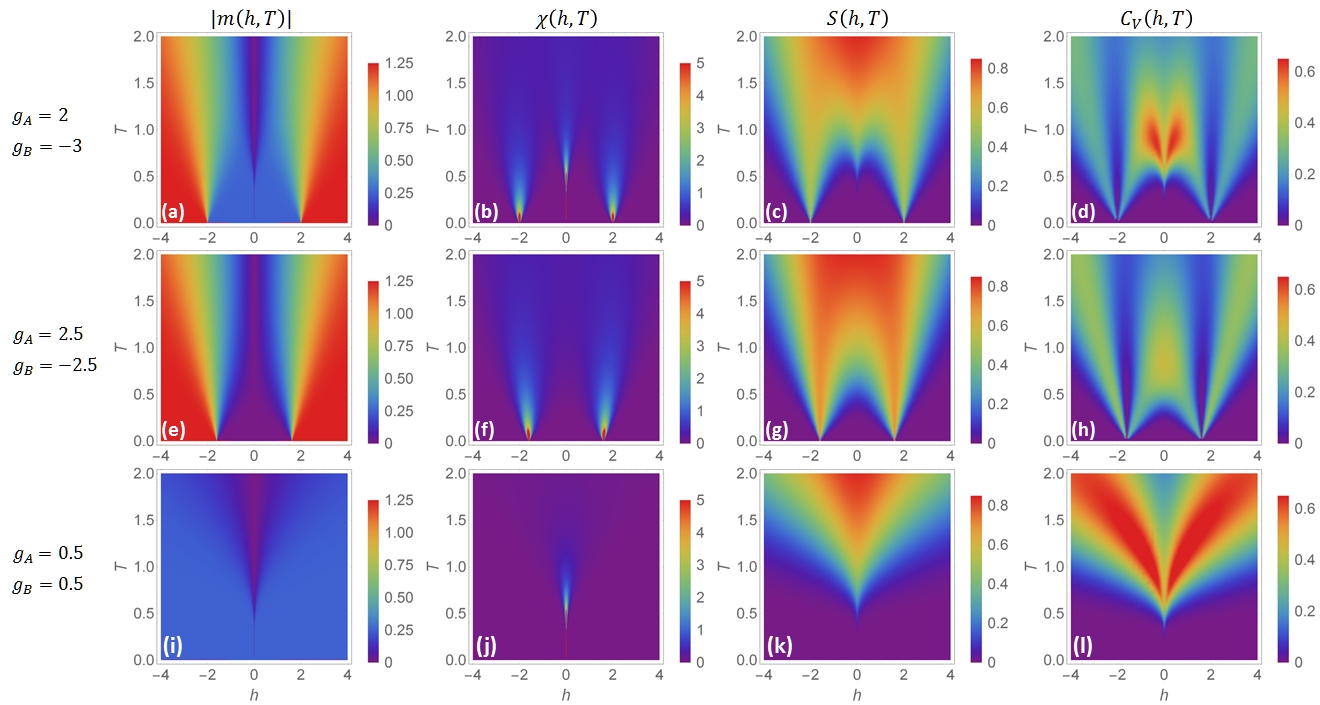}
        \caption{Color maps of the $T-h$ dependencies of the magnetization $|m(h,T)|$, the susceptibility $\chi(h,T)$, the entropy $S(h,T)/\ln2$, and the specific heat $C_V(h,T)$ for the nonuniform (the top row), the uniform staggered (middle row), and the uniform (the bottom row) $g$-factors.
        \label{hT}}
\end{figure*}

To explore the essential effects of nonuniform $g$-factors, we begin with the exactly solvable one-dimensional (1D) Ising model with alternating $g$-factors. This model captures the essential physics of the quasi-1D  Sr$_3$CuIrO$_6$ consisting of FM chains made up by alternating Cu and Ir ions [Fig.~\ref{geometry}(d)].  Our exact solution features the zero-field susceptibility with both FM and AF components, in agreement with the experiments.  It also turns out that the system possesses an exotic magnetic-field-driven critical point (CP) [Fig.~\ref{geometry}(e)] at which the spins with the smaller and larger magnitudes of $g$-factors are fully disordered and ordered, respectively, at absolute zero temperature---reminiscent of the all-time favorite aesthetic concept of `half fire, half ice.'  Since the Ising model 
is one of the main models of condensed matter and statistical physics \cite{Mattis2008} having been applied even to explain the activity of neurons in the brain \cite{Schneidman2006}, we anticipate that the present results to be widely applicable.

The model is defined by the Hamiltonian
\begin{equation}
\label{Ising}
H=-J\sum_{i=1}^N
{\sigma_i \sigma_{i+1}} - h\mu_\mathrm{B}S\sum_{i=1}^N{g_i \sigma_i},
\end{equation}
which corresponds to $N$ spins ($\sigma_i=\pm 1$) on a line with  periodic boundary conditions  $\sigma_{N+1}=\sigma_1$. $h$ is a uniform longitudinal magnetic field. The $g$-factors are alternating with $g_i=g_A $ for odd $i$ and $g_i=g_B$ for even $i$. We solve the model exactly using the transfer matrix method to calculate out the partition function and correlation functions (see Supplemental Material \cite{SI}).

\emph{The $T$-$h$ Phase Diagram---}The field $h$ on the nonuniform $g_A$ and $g_B$ factors can be regarded as the superposition of the uniform field of magnitude of $h(g_A+g_B)/2$ and the  staggered field of magnitude of $h(g_A-g_B)/2$ on a system with uniform $g\equiv 1$ factors. Therefore in Fig.~\ref{hT}, we contrast the $T$-$h$ dependencies of several thermodynamic quantities for the three cases: nonuniform $g_A=2$ and $g_B=-3$ (the top row), staggered $g_A=-g_B=2.5$ (the middle row, which is of the AF type), and uniform $g_A=g_B=0.5$ (the bottom row, which is of the FM type). Overall, the results for the $g_A=2$, $g_B=-3$ case are FM-like in the low-$h$, low-$T$ region. And they are AF-like in the high-$h$ region especially near the zero-temperature critical field $h_c$. Nevertheless, we notice two unusual features in the $g_A=2$, $g_B=-3$ case. First, the low-$h$, high-$T$ region is more AF-like. The other is that $h_c=2J/(\mu_\mathrm{B}Sg_A)$ depends on $g_A$, not on $g_B$. 
We elaborate the two anomalies below.

\begin{figure}[t]
        \includegraphics[width=\columnwidth,clip=true,angle=0]{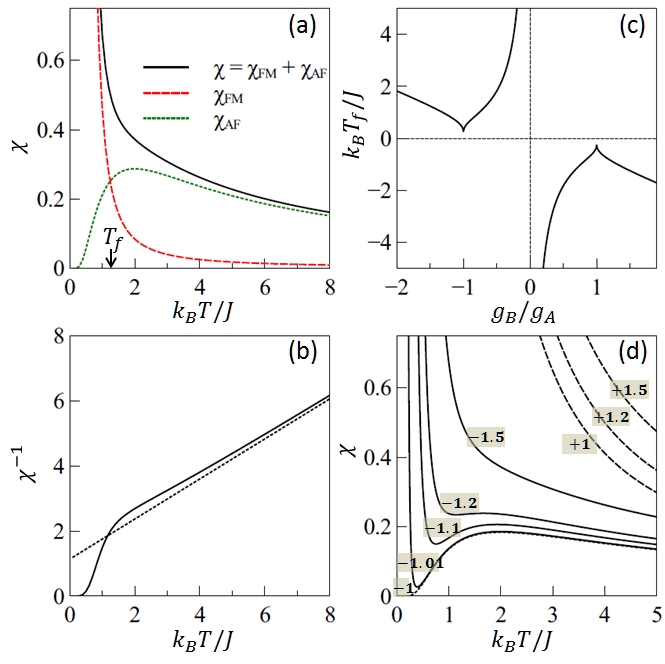}
        \caption{
        \textbf{The zero-field susceptibility} (a) $\chi(0,T)$ (the black solid line). Its FM-like component $\chi^{}_\mathrm{FM}$ (the red dashed line) and the AF-like component $\chi^{}_\mathrm{AF}$ (the green dotted line) cross at $T_f$. (b) The inverse susceptibility $\chi^{-1}$ (the solid line). Its high-$T$ Curie-Weiss behavior (the dotted line) wrongly suggests  that the exchange interaction is  AF. $J=1$, $g_A=1$, and $g_B=-1.5$. (c) $T_f$ as a function of $g_B/g_A$. (d) $\chi(0,T)$ for a variety of $g_B/g_A$ (the numbers with the shaded background).
       \label{fig_chi}}
\end{figure}

\emph{Zero-field susceptibility per site---}$\chi(h \to 0,T)$ takes the following elegant two-component form:
\begin{equation}
\label{chi}
\frac{\chi(0,T)}{\mu_\mathrm{B}^2S^2}=\beta e^{2\beta J} \Big(\frac{g_A+g_B}{2}\Big)^2 +\beta e^{-2\beta J} \Big(\frac{g_A-g_B}{2}\Big)^2.
\end{equation}
The coexistence of the two components requires that $g_A$ and $g_B$ have different magnitude: $|g_A| \neq |g_B|$. Having exponents of opposite sign, the two components are of the FM and AF types, respectively.  Eq.~(\ref{chi}) is equivalent to the combined magnetic susceptibility of the FM chain with the uniform $g=(g_A+g_B)/2$ factors and the AF chain with the uniform $g=(g_A-g_B)/2$ factors, which used to be called uniform and staggered susceptibilities, respectively \cite{Matsuura1976}. The results for the realistic values of $g_A=2$ and $g_B=-3$ in Sr$_3$CuIrO$_6$ are presented in Figs.~\ref{fig_chi}(a)(b). They agree qualitatively with the experimental results \cite{Niazi2002}, indicating that the FM Ising chain model with such nonuniform $g$ factors captures the essential physics of this copper-iridium oxide.

The spin frustration takes place when the two terms in Eq.~(\ref{chi}) are comparable. As shown in Fig.~\ref{fig_chi}(a), the FM term dominates at low $T$ and the AF one dominates at high $T$. The two terms become equal at
\begin{equation}
T_f=\frac{2J}{k_\mathrm{B}} \Big(\ln {\Big|\frac{g_A-g_B}{g_A+g_B}\Big|}\Big)^{-1}.
\end{equation}
The physical constraint of $T_f \ge 0$ requires [cf. Fig.~\ref{fig_chi}(c)]
\begin{equation}
\label{constraint}
J g_A g_B < 0 \;\;\; \mathrm{and}  \;\;\;  |g_A| \neq |g_B|.
\end{equation}
Otherwise, for $J g_A g_B > 0$, $T_f < 0$ is unphysical [Fig.~\ref{fig_chi}(c)]. Hence, to generate frustration in ferromagnets ($J>0$), the $g$-factors must have not only different magnitudes but also different signs.

To illustrate the effects of  the requirement in Eq.~(\ref{constraint}), we present $\chi(0,T)$ for a variety of $g_B/g_A$ \cite{Note:gAgB} while fixing $J=1$ in Fig.~\ref{fig_chi}(d). For the cases of $g_B/g_A<0$ (solid lines), $\chi(0,T)$ tends to dip down at an intermediate temperature near $T_f$, signaling the coexistence of comparable FM and AF components. On the other hand, for the cases of $g_B/g_A>0$ (dashed lines), the FM component dominates and frustration vanishes. One may get the intuition for the requirement as follows.  For FM $J>0$, $g_A$ and $g_B$ have opposite signs; thus, sufficiently strong $h$ will induce the AF alignment of the spins, interfering with the FM $J$. Nevertheless, it is striking that the spin frustration takes place even at $h \to 0$.

The generic method for distinguishing the exchange interactions is to characterize the high-temperature behavior of the magnetic susceptibility using the Curie-Weiss form of $\chi=C/(T-\theta)$ with $\theta > 0$, $=0$, and $<0$ corresponding to the FM, paramagnetic, and AF types, respectively. Indeed, as shown in Fig.~\ref{fig_chi}(b), the high-$T$ expansion of Eq.~(\ref{chi}) follows the Curie-Weiss law with
\ignore{
\begin{equation}
\chi^{}_\mathrm{HT} = \frac{C}{T-\theta},
\end{equation}
where
}
\begin{equation}
\theta=\frac{4J g_A g_B}{k_\mathrm{B}({g_A}^2+{g_B}^2)}, \;\;\;\;\; C=\mu_\mathrm{B}^2S^2\frac{g_A^2+g_B^2}{2k_\mathrm{B}}.
\label{CW}
\end{equation}
The constraint $Jg_Ag_B<0$ in Eq.~(\ref{constraint}) yields $\theta < 0$, which suggests the existence of AF exchange interaction \cite{Niazi2002}. However,  this is a pure illusion.

\emph{``Half Fire, Half Ice'' CP---}Furthermore, we found that the zero-temperature CP at $h_c$ in the present case of nonuniform $g$-factors is qualitatively different from that in the usual case of $g_A=-g_B$. We analyze the $T=0$ properties of the model Eq.~(\ref{Ising}) as a function of $h$. The magnetization per site is given by
\begin{eqnarray}
\frac{m(h,0)}{\mu_\mathrm{B}S}&=&\Big\{
       \begin{array}{ccc}
         (|g_B|-g_A)/2&  & 0<h<h_c \\
         |g_B|/2 & \mathrm{for} & h=h_c \\
         (|g_B| + g_A)/2 &  & h>h_c\\
       \end{array}
       \label{m}
\end{eqnarray}
At the critical field $h_c=2J/(\mu_\mathrm{B}Sg_A)$, the magnetization is $\mu_\mathrm{B}S|g_B/2|$, which is independent of $g_A$. This means that the spins on the $B$ sublattice are fully ordered, while the spins on the $A$ sublattice do not contribute to the magnetization. Moreover, the entropy per site reads
\begin{equation}
S(h_c,0)=\frac{1}{2}k_\mathrm{B}\ln 2.
\label{SA}
\end{equation}
Since the fully ordered spins on the $B$-sublattice contribute nothing to the entropy, Eq.~({\ref{SA}) means that the spins on the $A$-sublattice are fully disordered, which would be usually imaged to occur at infinite temperature. Thus, the spins are said to be hotter than fire on the $A$-sublattice but colder than ice on the $B$-sublattice, as artistically illustrated in Fig.~\ref{geometry}(e).

In comparison, the conventional Ising model with $J>0$ and $-g_B=g_A=g$ has equal contribution from the $A$- and $B$-sublattices by symmetry, yielding $m(h_c,0)/\mu_\mathrm{B}S=g/\sqrt{5} \approx 0.447\; g$ and $S(h_c,0)=k_\mathrm{B}\ln(\frac{1+\sqrt{5}}{2}) \approx 0.694\; k_\mathrm{B}\ln 2$, which are not the $-g_B=g_A$ limit of Eqs.~(\ref{m}) and (\ref{SA}). The entropy is $0.347\; k_\mathrm{B}\ln 2$ per sublattice, smaller than $0.5\; k_\mathrm{B}\ln 2$, which means the two sublattices are still correlated in short ranges at the CP. Indeed, the correlation function $\langle \sigma_{i} \sigma_{i+1} \rangle = 1-2/\sqrt{5}\approx 0.106$ at the CP for $-g_B=g_A$ but it is zero for $-g_B>g_A$ (see Supplementary Material \cite{SI}). Therefore, the `half-fire, half-ice' phase is identified as a new phase.

\begin{figure}[t]
       \includegraphics[width=\columnwidth,clip=true,angle=0]{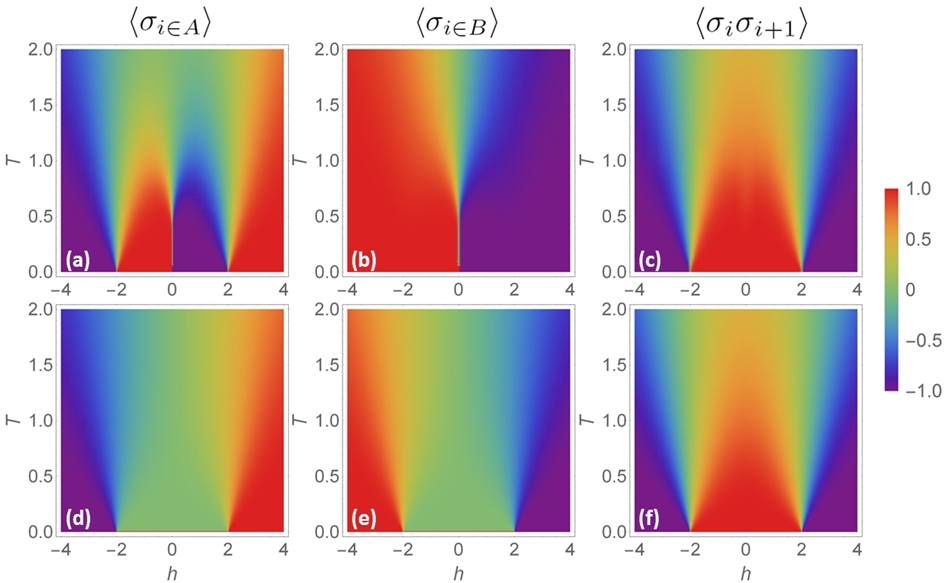}
        \caption{Color maps of the $T-h$ dependencies of the sublattice magnetization $\langle \sigma_{i \in A} \rangle$, $\langle \sigma_{i \in B} \rangle$, and the nearest-neighbor correlation function $\langle \sigma_{i} \sigma_{i+1} \rangle$ for the nonuniform $g$-factors $g_A=2$ and $g_B=-3$ (top row) and uniformly staggered $g$-factors $g_A=-g_B= 2$ (bottom row). $J=1$.
       \label{fig_corr}}
\end{figure}

More specifically, as shown in Fig.~\ref{fig_corr}, for strong field $|h|>h_c$, both the normal and the present cases have similar AF spin configuration of $\left|+-+-\cdots\right\rangle$ at low $T$, which gains energy from the $h$ term to overcome the energy cost from the $J$ term. On the other hand, for weak field $|h|<h_c$, the normal case's ground state is degenerate with the FM $\left|++++\cdots\right\rangle$ and $\left|----\cdots\right\rangle$ spin configurations, exhibiting $\langle \sigma_{i \in A} \rangle=\langle \sigma_{i \in B} \rangle=0$ and $\langle \sigma_{i} \sigma_{i+1} \rangle=1$, which gains energy from the $J$ term but nothing from the $h$ term. This degeneracy due to the $g_A=-g_B$ symmetry is lifted by $g_A<-g_B$, yielding the single spin configuration of $\left|----\cdots\right\rangle$, which gains energy from both the $J$ and $h$ terms. The above distinct behaviors for $|h|<h_c$ lead to the distinct CP's at $|h|=h_c$: It is now the normal case that gains energy from both the $J$ and $h$ terms with $\langle \sigma_{i \in A} \rangle=-\langle \sigma_{i \in B} \rangle=1/\sqrt{5}$ and $\langle \sigma_{i} \sigma_{i+1} \rangle \approx 0.106$, while the present case gains energy from the $h$ term but nothing from the $J$ term. Fig.~\ref{fig_corr} (top panel) also clearly shows that the `half fire, half ice' state survives a wide temperature range near $h_c$.

Being a highly degenerate state the `half fire, half ice' CP strongly responds to perturbations, such as the transverse exchange interaction $-J_\perp\sum_{i}
{(S_i^x S_{i+1}^x + S_i^y S_{i+1}^y)}$ known to be present in {\material} \cite{Yin2013}. At the CP, the $B$-sublattice spins are frozen in one direction  and the $A$-sublattice spins are ``paramagnetic.'' Thus, the $A$-sublattice spins can interact with each other via the virtual processes that flip the $B$-sublattice spins.
Hence the vicinity of the CP is described by the effective XY model
\begin{equation}
H_{XY}=\sum_{i\in A}
{\left[-J_\mathrm{eff}\left(S_i^x S_{i+2}^x + S_i^y S_{i+2}^y\right) - h_\mathrm{eff} S^z_i \right]}, \label{XY}
\end{equation}
where $J_\mathrm{eff}=J^2_\perp/2\Delta$ and $h_\mathrm{eff}=(h-h_c)g_A \mu_\mathrm{B}+J^2_\perp/4\Delta$ with $\Delta=|2h_c\mu_\mathrm{B}g_BS|$ being the energy cost to flip one $B$-sublattice spin. This model is exactly solvable by means of Jordan-Wigner transformation which reduces it to the model of noninteracting fermions with the spectrum $\epsilon(k) = |h_{\mathrm{eff}}|-J_\mathrm{eff}\cos k$. Hence it remains quantum critical for $|h_{\mathrm{eff}}| <  J_\mathrm{eff}$. The complete $J_\perp-h$ phase diagram as well as the responses to other stimuli such as the transverse field will be published elsewhere. 

\emph{Discussion.---}Nonuniform $g$-factors have been of interest in the studies of magnetic crystals \cite{Matsuura1976,Perk1975,Oshikawa1997} and clusters \cite{Ohanyan2015}, predominantly with pure $3d$ or pure $5d$ magnetic ions. The present exploration of the parameter space of $Jg_Ag_B<0$ and $|g_A|\ne|g_B|$ is motivated by studying the mixed $3d-5d$ materials \cite{Yin2013}. In particular, to see how unusual it is to reach the opposite signed $g_A$ and $g_B$ for FM $J>0$, $g_A=g_B$ was assumed in previous studies of Sr$_3$CuIrO$_6$ \cite{Nguyen1996,Furusaki1994}. We anticipate the understanding presented here to be widely relevant to other mixed $3d-5d$ transition-metal compounds, e.g., the `partial fire, partial ice' CP at which the spins with larger and smaller $|g_i|$ are frozen and boiling, respectively, in the other copper-iridium oxides \cite{Niazi2002,Battle1997,Battle1998,Blake1998}.

Last but not least, the sign of $J$ is not a prerequisite in the above analysis. With the constraint of $Jg_Ag_B<0$, it is equally acceptable to have AF $J<0$ and $g_A g_B > 0$. This is achieved by the transformation $\sigma_i \to -\sigma_i$ for sites in the $B$-sublattice only and the substitutions $J \to -J$ and $g_B \to - g_B$ in Eq.~(\ref{Ising}). Then, the AF exchange interaction implied by the high-$T$ Curie-Weiss behavior is real, while the low-$T$ response is FM-like. When the magnitudes of the $g$-factors are similar [c.f. the curve for $g_B/g_A=-1.01\%$ or $-1.1\%$ in Fig.~\ref{fig_chi}(d)], $\chi(0,T)$ looks like the susceptibility of many antiferromagnets, which goes down as $T$ decreases from the N\'{e}el temperature but mysteriously jumps up at very low $T$. The jump used to be attributed to the presence of magnetic impurity; the above results now offer an alternative mechanism. Most importantly, the analogous solution with AF exchange interaction greatly enlarges the range of materials containing the nonuniform $g$-factor induced frustration.

In summary, we have revealed novel magnetic frustration and an exotic `half fire, half ice' critical point in the Ising model with nonuniform $g$-factors. We have shown that these unusual $g$-factor effects are essential to understanding the mixed copper-iridium oxides. 
Considering the wide range of applications of the Ising model, we anticipate to realize the frustration and `partial fire, partial ice' states in lattice gas and neuron systems as well.

This work was supported by the U.S. Department of Energy (DOE), Office of Science, under Contract No. DE-AC02-98CH10886. C.R.R. participated in the research via Brookhaven National Laboratory Supplemental Undergraduate Research Program.
We are grateful to Tiffany Bowman for helping illustrate Fig.~1(e).


\begin{thebibliography}{20}
\bibitem [{Not()}]{Note:gAgB}%
  \BibitemOpen
  \href@noop {} {}\bibinfo {note} {It is adequate to discuss the case of $|g_B|
  \ge |g_A|$, since for $|g_B| \le |g_A|$ one simply swaps the $A$ and $B$
  labels to return to $|g_B| \ge |g_A|$.}\BibitemShut {Stop}%
\end{thebibliography}
%

\end{document}